\documentstyle[12pt,prl,aps,graphicx]{revtex}

\newcommand{\be}{\begin{equation}}
\newcommand{\ee}{\end{equation}}
\newcommand{\al}{\alpha}
\begin{document} \draft \date{\today} 
\title{"Non-$1/r$" Newtonian Gravitation and Stellar Structure} 
\author{Zakir F. Seidov\\
\it {Department of Physics, Ben-Gurion University of the Negev,
Beer-Sheva 84105, Israel}} \maketitle \begin{abstract}
Some analytical solutions for spherical-symmetrical equilibrium
configurations (planets/stars) in Newtonian Theory of Gravitation (NTG)
with deviations from $1/r$ law are discussed. 
Stability of star against the first-order phase transition is particularly
influenced by  deviations of degree in power-law potential .
\end{abstract}
\section{Introduction}
In the classical NTG, with the point-mass potential 
\be U(r) = G m / r, \ee
the equation of the spherical-symmetric equilibrium states reads:
\be {1 \over \rho(r)}{dP(r) \over dr} = - {G m(r) \over r^2}. \ee
Here, as usual, $G$ stands for constant of NTG, $P(r)$, $\rho(r)$ are
local pressure and density of stellar matter, at distance $r$ from the
 center of star,
and $m(r)$ is mass inside the sphere of radius $r$ ("$r$-sphere"). 
Provided EOS $P(\rho)$ is known, the Eq. (2) is ODE and is
degenerate in the sense, that any 
deviation from standard law (1) spoils the fundamental characteristic of 
spherical-symmetric stars - dependence of local gravitational force, at
radius $r$,
 only on the mass inside "$r$-sphere". That is any deviation of $1/r$ law
leads, in principle, to the enormous complexity of problem of the 
spherically-symmetric equilibrium states. Instead of
ODE one has to deal with integro-differential equation (IDE), see next
section.\\
In [1] authors discuss some recent schemes  of compactivation
leading to the generalized form of the
point-mass potential,
\be U(r) = {G m \over r} (1 + \alpha \  e^{-r/\lambda}), \ee
where $\alpha$ and $\lambda$
are constants. In the celestial mechanics, the various
"non-Newtonian" or "non-gravitational" forces have been considered
quite a long ago [2].\\
In this note I consider only the problem of stellar structure
in NNG (Non-Newtonian Gravitation). As nobody knows the $exact$ form of
"non-Newtonian" law, it is
reasonable to consider first the various simple cases of potential
$\,U(r)=Gm*f(r)\,$. Here I consider "power-law case", 
\be f(r) = 1\  / \ r ^{1 + \alpha},\ee where $\alpha$ is constant.
Some manifestations of this law in stellar structure are rather dramatic
that may help to judge in favor or against it. Despite this
"applicational" aspect it's certainly worth by itself considering the 
problem of stellar structure with non-$1/r$ gravitational
potential.\\
At first, I present the hydrostatic spherical-symmetric equilibrium
equation in general case.
\section{Equations of Stellar Structure in NNG}
If $f(x)$ is "generalized Newtonian" law, then the potential of thin
spherical shell is as follows: \be
u_{shell}(r,a)\ da=2\ \pi\ G\ \rho(a)\
a^2 \ da\ \int_{|r-a|}^{a+r}{f(x)x\over r\ a}dx.\ee
Here, $a$ and $da$ are radius and "thickness" of infinitesimally
thin shell, $\rho(a)$ is matter density at radius $a$, and $r$ is
distance from the center of spherical shell.\\
Note, that  only at $f(x)=1/x^1$ (and, of course at
$f(x)=1/x^0$, when point-mass potential itself
is constant!), the potential
inside the spherical shell is constant (and gravitational force is
zero). It's worth mentioning that both laws (3) and (4) give
the first-order correction leading to necessity of IDE, 
 while the  zeroth-order term coincides with $1/r$ law.\\
The potential $U_ {sphere}(\rho,\ r,\ R)$ , at distance $r$,  
of a sphere of radius $R$ is :
\be U_{sphere}(\rho,\ r,\ R)=\int_{0}^R u_{shell}(r,a)da.\ee
The Eq. (6) is valid both inside  and outside the sphere.\\
The equation of the hydrostatic equilibrium of
spherical-symmetric star reads:
\be {1\over \rho(r)}{dP(r) \over dr} =  {dU_{sphere}(\rho(r),\ r,\ R)
\over dr}.\ee 
The right side of this equation is  a double integral, and in
general is of much more complicated form than in the standard $f(r)=1/r$
case, Eq. (2). Provided EOS $P(\rho)$ is known, Eq. (7) is 
IDE for density function $\rho(r)$.\\
As at any arbitrary EOS $P(\rho)$ and non-$1/r$ law 
it's apparently impossible to solve the problem (5-7) , it's a good idea to
consider some simple cases first. In the next section, I
present the spherically-symmetric model
of a homogeneous imcompressible liquid, that corresponds to a polytropic
index $n=0$ in the polytropic EOS of the form $P(\rho)=K\ \rho^{1+n}$.
\section{Homogeneous sphere}
At the point-mass potential of form (4),  the gravitational
potential of homogeneous sphere
of radius $R$ and density $\rho=const$ is as follows:
$$aN \equiv N-\al;\ \ t \equiv r/R;$$
\be U_{int, out}= {2\pi G \rho R ^{a2}\over a1}
\left[ {(1+t)^{a3}-(\pm 1 \mp t)^{a3} \over a3\ t}-{(1+t)^{a2}\pm 
(\pm 1 \mp t)^{a2}\over a2}\right]. \ee 
Here upper and lower signs correspond to $t<1$ and $t>1$ respectively. 
Values of potential at the center and the surface of the homogeneous
sphere are: \be
U(t=0)=  4\pi G \rho R^{a2}/a2;\ \  
U(t=1)=  2\pi G \rho (2R)^{a2}/(a2\,a3).\ee 
The total gravitational potential energy of homogeneous sphere is:
\be  W= - \pi^2 G \rho^2 (2 R)^{a5}/(a2\,a3\,a5).\ee
Solution of equation (7) in this case leads to the following relation
for central pressure:
\be P_c=\rho*[U(t=0)-U(t=1)]=2\pi G \rho^2 R^{a2}(2-2^{a2}/a3)/a2.\ee
\begin{figure} \includegraphics[scale=.6]{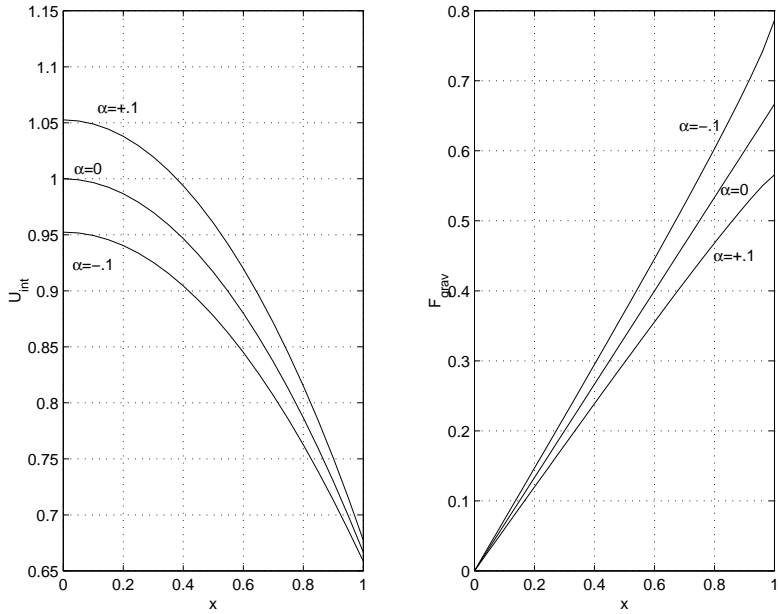}
\caption{Dependence of the potential (left panel) and the gravitational force
(right panel) inside the homogeneous
sphere for $1/r^{1+\al}$ point-mass potential. Vertical scales are somewhat
arbitrary as for different values of $\al$ dimensionality of constants are
different, see Eqs. (8-11).} \end{figure}
In Fig. 1,  some relevant curves are presented. 
Of course, in "real" cases it appears that $|\alpha|<<1$, so the Fig. 1
is presented here only for pure illustrative purposes. We  note that
nothing extraordinary interesting happens with the gravitational potential
of homogeneous sphere by broken $1/r$-law even for such large values of
$|\alpha|$. Suprisingly enough, even a such particular characteristic
of NTG as maximal force at surface of sphere survives in NNG.\\
It may be noted, that in spite of "enormous", in principle,  complexity of
the spherical-symmetric equilibrium equations
in NNG, in reality it is difficult to point out  any macro-effect
of "micro"-values of $\alpha$.\\
By the way, we notice a rather common opinion that in astronomical
("cosmic") scales, even small deviations from $1/r$ law may lead to some
macro effects. For example, external parts (outside the "$R$-sphere") of
homogeneous dark-matter or any other background matter are usually assumed
as non-influencing the equilibrium  and stability of spherical structure
with given radius $R$ [3].\\ 
It's seems worth reconsidering this problem with account of deviations
from $1/r$ law. Two potentially important cases may be the models with
adiabatic index of EOS close to 4/3, suggested  sometimes as model of 
supermassive stars (quasars), and the isotermic configurations
suggested as model of inter-(or proto-) stellar clouds or other cosmic
structures [9].\\
Here I consider one rather interesting effect of
NNG. In NTG (as well as in GR) the problem of
equilibrium and stability of a star (planet) with first-order  phase
transition (PT1) exists apparently since pioneer works by Ramsey [4].\\
 In "classical" NTG, the elegant and quite suprising result is that if PT1
occurs at the star's center with $q>3/2$, $q=\rho_2 / \rho_1$ being the
ratio of new-to-old phase densities, the star undergoes a stability loss
irrespective of EOS! \\
This "exact" value, $3/2$, has apparently its origin
from another two exact values: 1, power of $r$ in point-mass potential
law, and 3, dimension of space. It's the place here to illustrate the
first part of this assertion, for the simple model. \\
The simplest model of star (planet) with PT1 is a
two-zone model with constant densities $\rho_1$ and $\rho_2$ in envelope
and core respectively. 
\section{Two-zone model with PT1 }
In this model density distribution is a piece-wise function:
$\rho(0<r<r_n)=\rho_2=const$, and $\rho(R>r>r_n)=\rho_1=const$,
with $q=\rho_2\ / \rho_1\  >1$; here $r_n$ and $R$ are radii of core  and
of total star. At the core-envelope boundary, the pressure is constant,
$P(r_n)=P_0=const$. The potential in
envelope may be considered as a sum of $internal$ potential of homogeneous
sphere of density $\rho_1$ and radius $R$ and of $external$ potential
of homogeneous sphere of density $\rho_2 - \rho_1$ and radius $r_n$:
\be 
U_{env}(\rho_1,r,R)=U_{int,sphere}
(\rho_1,r,R)+U_{out,sphere}(\rho_2-\rho_1,r,r_n),\ee 
where two functions in the right site are given in Eq. (8).
Hydrostatic equilibrium equation (7) has the solution, in the envelope,
$r>r_n$:
\be P(r)=\rho_1\ [U_{env}(\rho_1,r,R)-\ U_{env}(\rho_1,R,R)];\ee
\be P_0=\rho_1\ [U_{env}(\rho_1,r_n,R)-\ U_{env}(\rho_1,R,R)].\ee
The last expression is "initial condition" 
for the hydrostatic equilibrium equation (7) and determines the total
radius $R$ as function of radius of core $r_n$.
Total mass of model is:
\be M=4\ \pi\ G\ [\rho_1\ R^3\ +(\rho_2-\rho_1)r_n^3]/3. \ee
On the other hand, $P_0$ is central pressure for "initial" 
homogeneous configuration with "initial values" of total mass $M_0$,
total radius $R_0$, which are defined as in Eq. (11) with $R=R_0$:
\be P_0=2\pi G \rho_1^2 R_0^{a2}(2-2^{a2}/a3)/a2;
M_0=4\ \pi\ G\ \rho_1\ R_0^3\ /3.  \ee
These relations allow to exclude $P_0$ from Eqs. (13-15) and  to express 
$R$ and $r_n$ in units of $R_0$, and also $M$ in units of $M_0$.\\
 According to the static criterion [5], 
the dependence of a total mass of equilibrium states on central pressure 
determines the stability of models.
That is the branch of $M_{eq}(P_c)$ with positive derivative $dM_{eq}/dP_c$ presents 
the stable equilibrium states while the branch with negative 
$dM_{eq}/dP_c$ presents unstable equilibrium states.\\
In our case we can use $r_n$ (or even $r_n/R$) as independent variables
as they both are monotonic functions of $P_c$.
\subsection{Classic NTG Case}
For convenience, I present here, very briefly, some relevant formulae (and
results) in NTG. Although they are known since quite a long ago [3] they
have been rederived in literature sometimes, e.g. [6].\\
Instead of (13-16) we have (all variables are expressed in relevant
"initial" values):
\be M=((q-1)x^3)+1)R^3,\ \ \  x=r_n/R;\ee
\be R^3-[1-(2q-3)r_n^2]R-2(q-1)r_n^3=0;\ee
\be R=[1+(2q-3)x^2-2(q-1)x^3]^{-1/2};\ee
These equations allow a full analytical treatment of
the problem. In particular:\\
 $dM(x)/dx$ may be negative only at $q>3/2$;\\
 $dM(x)/dx=0$ at $x$ determined by the equation \\
$(q-1)^2x^4+4(q-1)x-(2q-3)=0$;\\
the last equation has solutions $x<\sqrt 2 -1$;\\
at mass minima, $M=R^2$.
\subsection{The $1/r^{1+\al}$ Case}
As relevant formulae are rather cumbersome in general case  I'll not 
rewrite them from Eqs. (8-16), and only consider a situation with small
cores. 
For small values of a relative radius of core, $x=r_n/R<<1$, we have
expansion up to $x^3$:
\be R=\left (1+
{(3-\al)(2-\al) \over\ 6-2\al -2^{2-\al}}
\left[ {2^{2-\al}(q-1)\over (2-\al)(3-\al)}x^{2-\al}
-{(1-\al)\over 3}x^2 
-{2(q-1)\over 3}x^3 \right]\right)^{-1/(2-\al)}.\ee
This equation is valid for arbitrary $\al$ and for small $x$.
At the limit $\al \rightarrow 0$ this equation coincides  with exact
 expression (19) which is valid for $any$ $x$, in NTG case, 
providing $q_{cri}=3/2$.\\
 However the case of non-zero $\al$'s  differs from the
classic case qualitatively. That is we don't have,
at small values of $\al$, some small correction to 
critical value of PT1 in NTG $q_{cr}=3/2$. 
Instead, at $\al>0$ ,  the smallest power of $x$ in Eq. (20) is
$a2\,<\,2$, the corresponding coefficient is negative for any $q>1$,
therefore we have $dR/dx<0$. As evident from (15), in $M$ "additional"
term $\propto x^3$, that is $\Delta M \propto \Delta R \propto
(x^{2-\al}\ \mbox{or}\ x^2)$, that provides the continuity of
derivative $dM/dR$ along the curve of equilibrium states [7]. \\
In other words, at $\al>0$ (more steeper  dependence of potential as compared
with $1/r$ law of NTG), we have $q_{cr}=1$,
that is for small cores there is a situation of the $absolute\ 
 instability$ of star against  PT1; while at $\al<0$, there is reverse
situation of the $absolute\ stability$ of star against PT1.\\ 
For the larger (but still small enough) dimensions of core there takes
place a restoration of the normal behavior of $M(x)$ curves, that is,
$dM/dx>0$ at $q<3/2$ and $dM/dx<0$ at $q>3/2$, see Fig. 2. \\ 
\begin{figure}
\includegraphics[scale=.4]{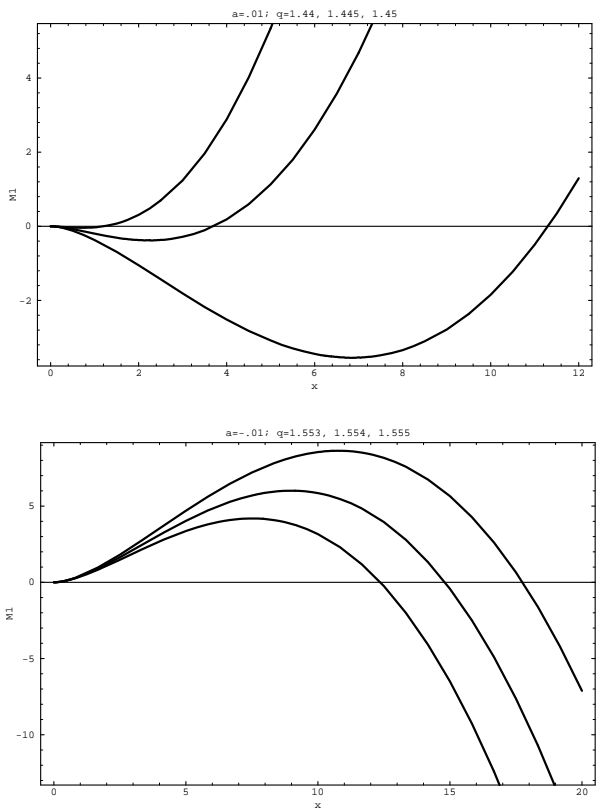}
\caption{Dependence of the total mass of star with first-order phase transition 
on the relative radius of new-phase core, in NNG. Abscissae are 
$10^6\,r_n/R$, ordinates are $10^{13}\,(M-1)$. The upper panel
 corresponds to the case of positive $\al =.01$. Curves (starting from upper one) are
shown for three values of  $q=1.44,\,1.445\,\mbox{and} \,1.45$, respectively. 
Note, that,  in principle, at $\al >0$, all curves at $x=0$ point have
negative derivative,
that is there is a minor region of small-core instability  for any $q>1$.\\
The reverse situation of negative $\al =-.01$ is shown in the lower panel. 
Three curves (starting from upper one) correspond to values 
of $q=1.555,\,1.554\,\mbox{and}\,1.553$. 
Here, all curves at $x=0$ point have positive derivative, that is there is
a minor region of small-core stability  for any $q>1$.} \end{figure}
The region of "abnormal" behavior of $dM/dx$ may be  very small,
a value of $x$, at which again $R=1$ and $M=1$ is:
\be x1=\left ( {3\,2^{a2}\,(q-1)\over (2-\al)(3-\al)(1+\al)}\right
)^{1/\al}.\ee. 
It's worth mentioning that $x1$ is very sensitive function of $q$ at
small enough values of $\al .$  It seems that despite the
smallness of $real$ values of $\al$, and therefore the smallness of the
"abnormal" stability region, this influence of the power-law potential
 on stabilty of a star is potentially of some major interest.
\section{Conclusion}
The deviations from $1/r$ law of NTG revealing itself in changing the 
value power of $r$
lead to a dramatic change of the stellar stabilty against the
first-order
phase-transition in the center of star, with small new-phase cores. 
The  steeper dependence of point-mass potential on $r$ leads
to the $absolute\ instability$ of star at smaller core dimensions, while
the lesser (than reciprocal) dependence of the point-mass potential leads
to the "absolute stability" of star. Though for "real"  values of $\al$
this phenomenon takes place only at the very small dimensions of new-phase
cores, nevertheless it may, in principle,  serve as some means 
for proving or rejecting such form of deviations from $1/r$-law. One may
note, for example, the  microcollapses of neutron star
revealing itself in pulsar timing "faults" [8].
Another possibility may occur in the larger astronomic
structures, such as IMC [9] and supermassive stars [3, 5]. 

\end{document}